\documentclass[a4paper,fleqn,usenatbib]{mnras}

\usepackage{newtxtext,newtxmath}

\usepackage[T1]{fontenc}
\usepackage{ae,aecompl}
\usepackage[dvipsnames]{xcolor}

\usepackage{graphicx}	
\usepackage{amsmath}	
\usepackage{amssymb}	

\title[Chemically peculiar ejecta in $\eta$ Carina]{The peculiar chemistry of the inner ejecta of Eta Carina}

\author[C. Bordiu et al.]{
Cristobal Bordiu,$^{1}$\thanks{E-mail: cbordiu@cab.inta-csic.es}
J. Ricardo Rizzo$^{1, 2}$
\\
$^{1}$Centro de Astrobiolog\'ia (INTA-CSIC), Ctra. M-108, km. 4, 28850 Torrej\'on de Ardoz, Madrid, Spain\\
$^{2}$ISDEFE, Beatriz de Bobadilla, 3, 28040 Madrid, Spain\\
}

\date{Accepted XXX. Received YYY; in original form ZZZ}

\pubyear{2019}

\begin{document}
\label{firstpage}
\pagerange{\pageref{firstpage}--\pageref{lastpage}}
\maketitle

\begin{abstract}
We investigated continuum and molecular line emission of four species (CO, HCN, H$^{13}$CN, and HCO$^+$) at 0.8 mm in the inner region around $\eta$ Car, using ALMA archival observations at a resolution better than 0.2 arcsec. We report the discovery of an asymmetric extended structure northwest of the star, independent from the continuum point source. The structure is only traced by continuum and HCO$^+$, and not detected in the other lines. 
Kinematics of this structure reveal that the HCO$^+$ gas likely arises from ejecta expelled in the 1890s eruption. The ejecta is propagating outward within the cavity produced by the current wind-wind interaction of $\eta$ Car A and its companion. Chemical analysis of the ejecta reveals an apparent lack of CO and nitrogen-bearing species. We explore possible explanations for this peculiar chemistry, that differentiates this structure from the ejecta of the Great Eruption, rich in HCN and H$^{13}$CN. We also report an absorption component near the continuum point source, only traced by HCN and H$^{13}$CN in their vibrational-ground and vibrationally-excited states. This absorbing gas is attributed to a hot bullet of N-enriched material expelled at a projected velocity of 40 km s$^{-1}$.
\end{abstract}




\section{Introduction} \label{sec:intro}

Luminous blue variables (LBVs) are evolved massive stars characterized by remarkable spectrophotometric variability and heavy mass loss in the form of dense and steady winds and occasional eruptions. LBVs shed out large amounts of dust and CNO-enriched material into their surroundings. The shocks and strong FUV fields produced by LBVs represent a continuous energy input to the stellar neighbourhood, eventually altering its structure and composition.

$\eta$ Car is the archetypal member of the LBV family, and one of the most luminous sources in the Galaxy ($5\times10^6\, \mathrm{L}_{\sun}$). It is a very eccentric binary system composed by a LBV star ($\eta$ Car A) of $\sim$ 100 $\mathrm{M}_{\sun}$ and a hotter, less massive companion ($\eta$ Car B) of about 30--40 $\mathrm{M}_{\sun}$ \citep{2001ApJ...553..837H, 2010ApJ...710..729M}. $\eta$ Car is located at 2350$\pm$50 pc in the Carina arm \citep{2006ApJ...644.1151S}, with a LSR velocity of $-$19.7 km s$^{-1}$  \citep{2004MNRAS.351L..15S}. Its peculiarity, brightness and closeness represent a unique chance to witness the last breaths of a high mass star. Consequently, $\eta$ Car and its surroundings have been exhaustively observed during the last decades, becoming one of the best studied objects in the Galaxy (e.g. \citealt{1995AJ....109.1784D}, \citealt{2003ApJ...586..432S},  \citealt{2008Natur.455..201S}).

$\eta$ Car underwent a devastating outburst in the 1840s, an event nicknamed the Great Eruption that expelled more than 40 $\mathrm{M}_{\sun}$ of processed material \citep{2010MNRAS.401L..48G, 2017ApJ...842...79M}, forming the bipolar Homunculus Nebula \citep{1989ASSL..157..101D}. This outburst was followed by a second, lesser eruption in the 1890s that gave birth to a smaller structure known as the Little Homunculus \citep{2003AJ....125.3222I}. Around the 1940s, a third abrupt rise in the light curve of $\eta$ Car happened, followed by a continuous luminosity increase which persists until the present days \citep{2009A&A...493.1093F}. This peculiar behaviour of the light curve has been explained as caused by the dissipation of obscuring debris in the line of sight \citep{2019MNRAS.484.1325D}.

The innermost region of $\eta$ Car is a hostile environment, subject to strong FUV/X-ray fields \citep{2002A&A...383..636P,2017ApJ...838...45C,2018NatAs...2..731H} and periodical shocks governed by the 5.54-year orbital cycle of the binary \citep{2000ApJ...528L.101D}. In particular, \cite{2008MNRAS.388L..39O}, \cite{2009MNRAS.396.1308G}, \cite{2013ApJ...773L..16T} and \cite{2016MNRAS.462.3196G} studied the complex wind-wind interactions of $\eta$ Car A and its companion, revealing multiple structures surrounding the binary. These structures, detected in forbidden emission lines of Fe and N, allowed constraints to the orbital geometry of the system and the properties of the stellar winds.

\cite{2016MNRAS.462.3196G} depicts a very complex scenario in which a number of substructures coexist at different velocity regimes and evolve throughout the orbital cycle depending on the ionization state: (1) a low-density cavity is carved out when the hot and fast wind of $\eta$ Car B disrupts the slower and denser primary wind. This structure may be linked to the multiple hydrogen radio recombination lines (RRLs) that were detected towards the NW quadrant of $\eta$ Car \citep{2003MNRAS.338..425D, 2014ApJ...791...95A}, which possibly arise from the ionized gas within the cavity; (2) multiple arc-like structures, visible in [\ion{Fe}{ii}] and [\ion{Fe}{iii}], trace expanding fossil shells of compressed, high-density primary wind that pile up with each 5.54-year cycle. These structures are formed due to the passage of $\eta$ Car B through the massive primary wind during periastron, when the two stars get as close as 1.5 au. The shells are particularly prominent in the far side of $\eta$ Car. Fainter [\ion{Fe}{ii}] arcs and clumps, likely the remnants of even older cycles, are still visible at larger distances from the star, moving at velocities comparable to the terminal velocity of the primary wind, of 420 km s$^{-1}$ \citep{2012MNRAS.423.1623G}; and (3), slowly-moving clumps of debris from the 19th century events dominate the emission in the near side of $\eta$ Car, propagating in the direction of apastron. This matter includes the Weigelt blobs and their associated extended structures, visible at infrared wavelengths \citep{2005A&A...435.1043C}.

The Weigelt blobs are close-in ejecta located at $\sim$0.1--0.3 arcsec of the star, moving at $\sim$ --40 km s$^{-1}$ near the orbital plane \citep{1986A&A...163L...5W}. These clumps are partially ionized and exhibit rapidly varying spectral features modulated by the orbital cycle (\citealt{2005A&A...436..945H, 2006A&A...452..253J}). The ejection of the blobs took place at some point between 1880 and 1930 \citep{2012ASSL..384..129W}. \cite{2016MNRAS.462.3196G} argues that the Weigelt blobs are just the ionized surfaces of a larger neutral structure, like `blisters' in the skin of a cloud of dust and gas, exposed to mid/far ultraviolet (MUV/FUV) radiation from the stars. Therefore, the slowly-moving material that propagates within the wind-blown cavity was most likely expelled during the 1890s eruption.

A significant fraction of the (sub-)millimetre continuum flux from $\eta$ Car corresponds to free-free radiation, whereas thermal dust dominates the infrared continuum \citep{1995A&A...297..168C}. A detailed study of the dust content of the Homunculus combining legacy \textit{ISO} data with \textit{Herschel} observations allowed \cite{2017ApJ...842...79M} to identify a primary source of thermal radiation in the central 5$\times$7 arcsec around $\eta$ Car. In addition, a separate feature at 350 $\mu$m that exhibited a certain level of variability was detected. This second feature was consistent with a compact source of $\sim2$ arcsec, being also tentatively linked to the free-free emitting region reported by \cite{2014ApJ...791...95A}. 

Molecular spectroscopy, in combination with the resolving power of interferometry, could provide a complementary point of view of the complex central region of $\eta$ Car. The study of the formation and survival of molecules and the evolution of its relative abundances is key to understand the physical and chemical processes involved in such a harsh environment. Pioneering research by \cite{2001ApJ...553L.181R, 2003A&A...411..465R, 2008ApJ...681..355R} demonstrated the potential of molecular gas associated with evolved massive stars as a tool to understand their evolution and mass-loss history. Molecular hydrogen around  $\eta$ Car was first detected by \cite{2002MNRAS.337.1252S}, distributed over the surface of the Homunculus. Later, \cite{2006ApJ...645L..41S} reported a tentative detection of NH$_3$ towards the inner region of the nebula --although \cite{2016ApJ...833...48L} claims this detection as arising from the H81$\beta$ RRL at 23.861 GHz--. \cite{2012ApJ...749L...4L} carried out the first molecular survey with APEX, reporting the detection of eight species --including CO and four N-bearing molecules such as CN, HCN, and HNC--. Recently, ALMA observations with a resolution of about 1 arcsec \citep{2018MNRAS.474.4988S} revealed an expanding torus of CO in the waist of the Homunculus. A similar structure has been already discovered around the LBV object MN101 (MGE042.0787+00.5084), possibly in a more advanced stage \citep{2019MNRAS.482.1651B}.

In this work we present  ALMA archival observations of CO, HCO$^{+}$, HCN and H$^{13}$CN towards $\eta$ Car with an unprecedented resolution better than 0.2 arcsec. We analyse the spatial distribution and kinematics of the emission in the innermost region of the Homunculus, also providing hints on its chemistry and linking the observed features to the violent history of $\eta$ Car.

\section{Data and overall results} \label{sec:results}

We make use of ALMA band 7 archival observations from project 2016.1.00585.S (P.I: G. Pech-Castillo). The source was observed on 2016 October 24 under excellent weather conditions --0.57 mm of precipitable water vapour--, with an integration time of 668 s. A total of 41 12-m antennas were used, providing a maximum baseline of 1.8 km. Quasars J1107-4449, J1047-6217 and J0538-4405 were used for flux, phase and bandpass calibration respectively. The correlator was set to observe four simultaneous spectral windows of 1 GHz each, targeting the rotational lines of CO $J=3\rightarrow2$ (345.795989 GHz), H$^{13}$CN $J=4\rightarrow3$ (345.339769 GHz), HCN $J=4\rightarrow3$ (354.505475 GHz) and HCO$^{+}$  $J=4\rightarrow3$ (356.734223 GHz). Visibilities were reduced following the standard ALMA pipeline with \textsc{casa} v.4.7.0 r38335. The resulting QA2 products included four spectral cubes and a continuum map, with a characteristic beam of 0.17$\times$0.13 arcsec (P.A. -60\degr). The phase centre is shifted by ($-0.\arcsec1$,$ +0.\arcsec15$) with respect to the continuum peak, presumably the star.

\begin{figure*}
\includegraphics[width=\textwidth]{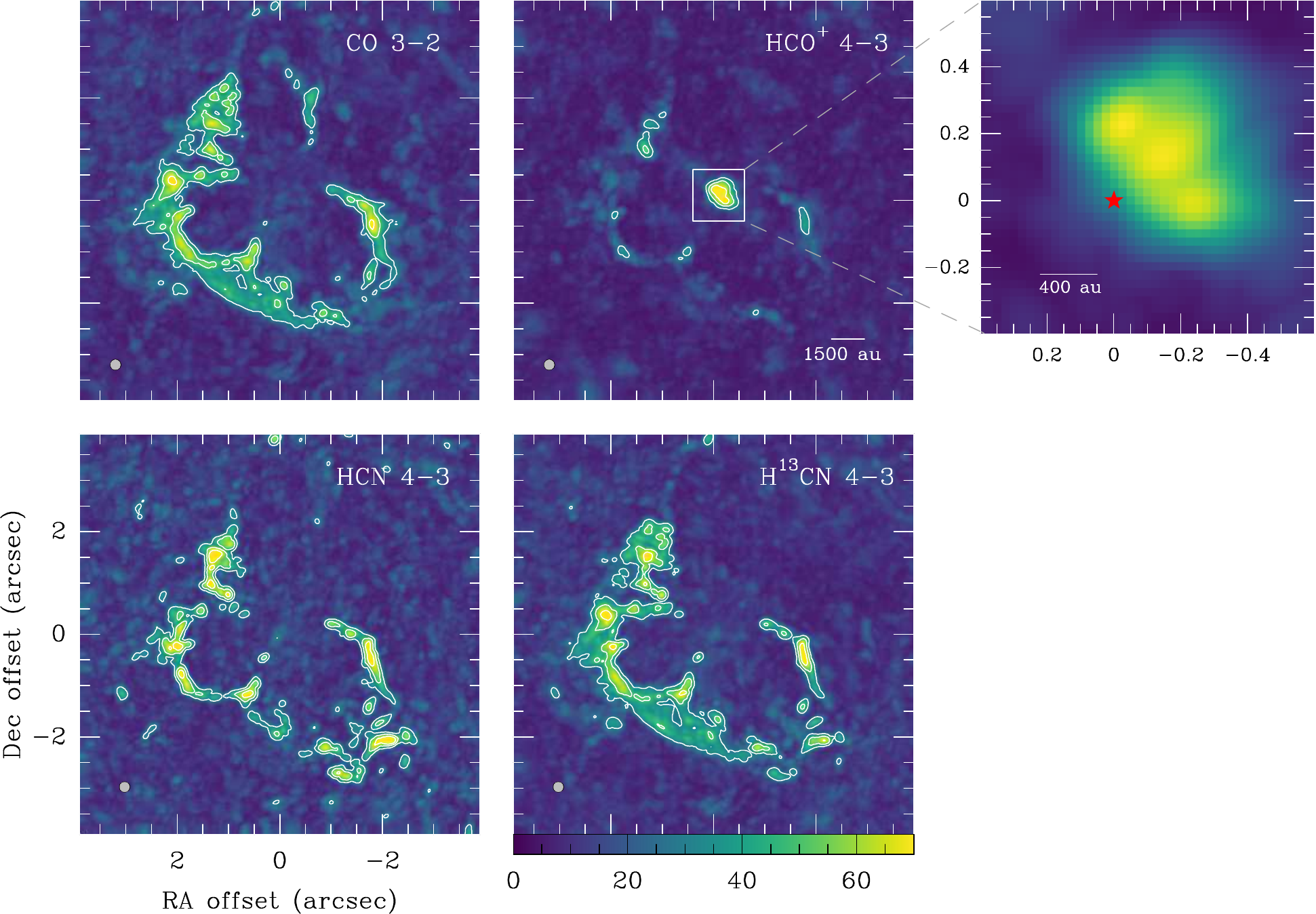}
\caption{Peak-intensity maps of CO $J=3\rightarrow2$, HCN $J=4\rightarrow3$, H$^{13}$CN $J=4\rightarrow3$ and HCO$^+$ $J=4\rightarrow3$ in colour scale. Spectral lines are indicated in the top right corner. Contours are 30, 50 and 70 K. A close-in view of the central HCO$^+$ emission (the Peanut) is shown in the inset, with the position of  $\eta$ Car indicated by the red marker. Beam width is shown in the bottom left corner of each panel.}
\label{fig:fig1}
\end{figure*}

\begin{figure*}
\includegraphics[width=\textwidth]{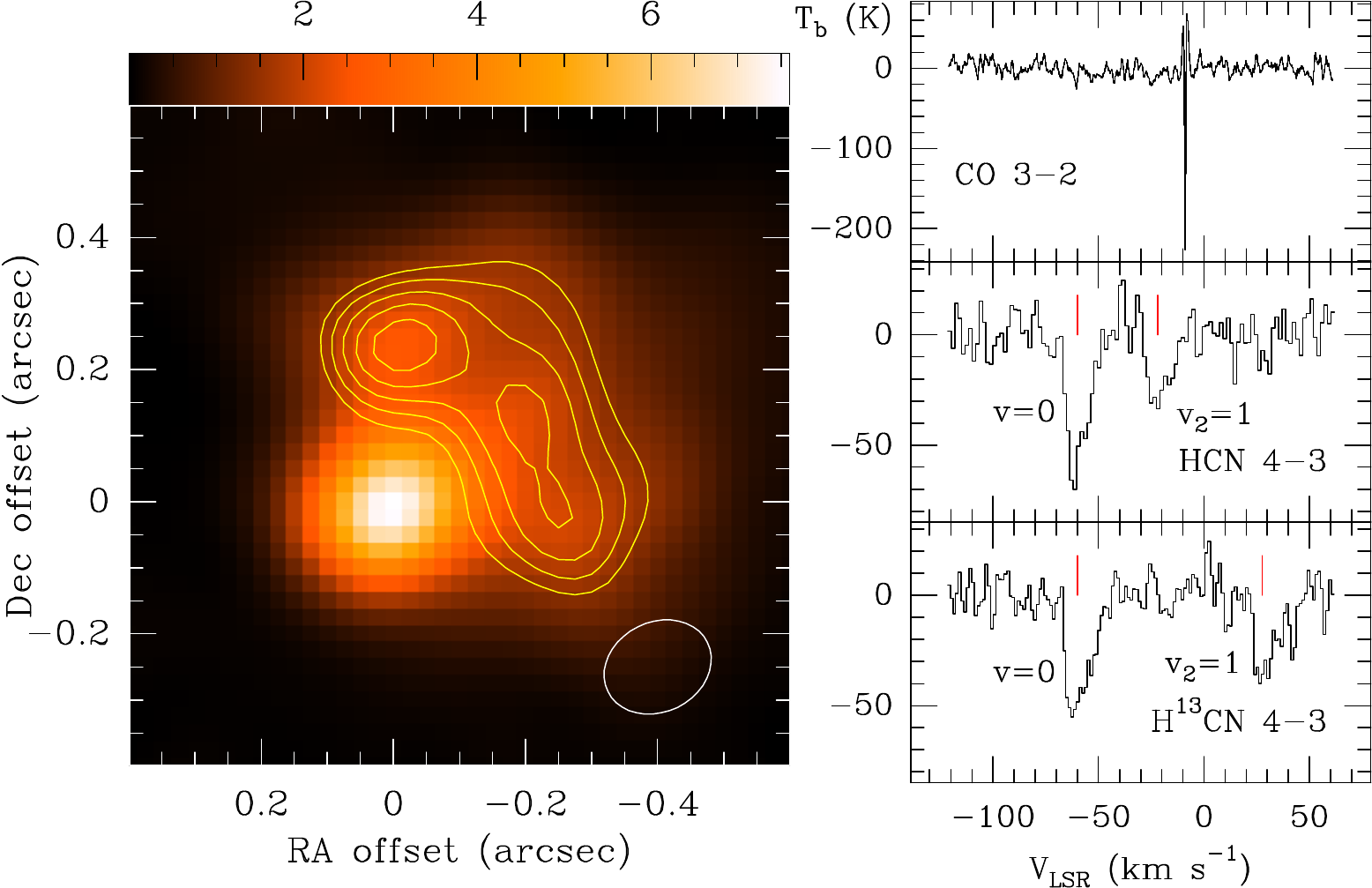}
\caption{Continuum emission and line absorptions at 345.8\,GHz. The continuum image (left, colour scale) is composed by a point-like source and an extended source coincident with the Peanut. Contours are 1.4, 1.6, 1.8, 2.0, and 2.2 Jy~beam$^{-1}$, and correspond to the resultant image after a point source subtraction. Half power beam width is indicated by the white ellipse at the lower right corner. Spectra depicted in the right panels are from an average of 0.1 arcsec around the point source. The CO line is at $-9$\,km\,s$^{-1}$, while the HCN and H$^{13}$CN lines  appear at $\sim -60$\,km\,s$^{-1}$ (red vertical marker).}
\label{fig:fig2}
\end{figure*}

We adopt the following conventions: (1) intensities are given in a brightness temperature scale ($T_\mathrm{b}$). The conversion from flux density to $T_\mathrm{b}$ is made through the expression

\begin{equation}
T_\mathrm{b} = 1.22\times10^6 \frac{S_\nu}{\nu^2\theta_\mathrm{maj}\theta_\mathrm{min}}
\label{eq:conversion}
\end{equation} 

\noindent with $T_\mathrm{b}$ in K, $S_\nu$ in Jy beam$^{-1}$, $\nu$ the frequency in GHz and $\theta_\mathrm{maj}$ and $\theta_\mathrm{min}$ the major and minor beam sizes in arcsec; (2) positions are offsets relative to the J2000 coordinates of the source;\footnote{$\alpha =10^\mathrm{h}45^\mathrm{m}03^\mathrm{s}.5362, \,\delta=-59^\circ 41\arcmin 04\arcsec.0534$ (J2000) \citep{2018yCat.1345....0G}.} and (3) velocities are expressed in the local standard of rest frame (LSR).

Fig. \ref{fig:fig1} presents the peak-intensity maps of CO, HCN, H$^{13}$CN and HCO$^{+}$ in a region of 8$\times$8 arcsec around the star. In CO, HCN and H$^{13}$CN we distinguish a clumpy C-shaped structure that surrounds the binary at an average radius of $\sim2$ arcsec --about 4700 au--. This structure corresponds to the disrupted torus described by \cite{2018MNRAS.474.4988S} from CO $J=2\rightarrow1$ observations, who dated it back to the Great Eruption. The torus traces the bright rims of the so-called `butterfly nebula', a region of efficient dust formation surrounding $\eta$ Car, clearly visible in mid-IR images \citep{2005A&A...435.1043C}. The clumpiness of the gas in the torus translates into multiple velocity components in the range ($-100, +100$) km s$^{-1}$.

Contrary to CO, HCN and H$^{13}$CN --which display a remarkably similar spatial distribution--, HCO$^+$ tells a very different story. The torus is still visible as faint spots, but the most intense emission, with a peak intensity of $77.5\pm6.8$ K, arises from a slightly elongated region very close to the star, roughly $0.6\times0.4$ arcsec. Hereafter we refer to this structure as `the Peanut' due to its particular shape (see inset).

Continuum emission is only detected at a significant level (5$\sigma$) in the inner $\sim 0.6$ arcsec, as shown in Fig. \ref{fig:fig2}. It depicts a point-like source --presumably related to $\eta$~Car-- and an extended component matching the Peanut. We attempted to isolate the point source emission by fitting it to a single 2D Gaussian. Surprisingly, the point source has an extension slightly larger than the beam, with a deconvolved size of $0.11 \times 0.11$ arcsec (i.e. 260~au). This unresolved region is fairly smaller than the estimates by \cite{2014ApJ...791...95A} for the hydrogen RRL emitting region. The contours in Fig. \ref{fig:fig2} show the extended component isolated from the point source. The peak intensity of the point source is $6.41 \pm 0.02$~Jy~beam$^{-1}$, while the integrated fluxes of the point-like and extended sources are $13.1\pm 0.5$ and $17.4\pm 1.1$~Jy, respectively. This flux is remarkably below those quoted by \cite{2014ApJ...791...95A} at 225.4 and 291.2 GHz, even when adding up the point source and the extended component (30.5 Jy). This is particularly surprising as observations were gathered close to apastron, when flux is expected to be maximum \citep{2005ASPC..332..126W}. However we note that these fluxes correspond to apertures of about $\sim$ 1 arcsec$^2$. We do not observe significant emission at this scale, but integrating a circular area of this size yields a flux of  $38.9 \pm 0.6$~Jy, which is in very good agreement with previous results. This reveals the existence of a low-intensity plateau at the noise level that surrounds the resolved structures.

The right panels of Fig. \ref{fig:fig2} also show the discovery of CO, HCN and H$^{13}$CN absorbing the continuum. The CO absorption is projected onto the whole continuum-emitting region, with a central velocity of $-9$ km~s$^{-1}$ and a width of $\sim 0.6$ km~s$^{-1}$. The absorptions of HCN and H$^{13}$CN are notoriously different: they are point-like, located very close to the star and centred at $-60$~km~s$^{-1}$, with widths of $\sim 12$ km~s$^{-1}$. Strikingly, we also detected the $v_2=1$ lines of HCN and H$^{13}$CN, as shown in the figure. We analyse the distribution and excitation of these lines in Sect. \ref{sec:disc-abs}.

\section{The Peanut}\label{sec:disc-pea}

The Peanut is the most prominent emission feature in the central region. It is located to the NW of $\eta$ Car, at a projected distance of just $\sim 0.1-0.2$ arcsec (<1000 au) from the star position. Its projected major axis is aligned in the NE-SW direction (see Fig. \ref{fig:fig1}). To the best of our knowledge, this is the very first direct detection of this structure beyond infrared; but considering the size uncertainty, it may well be related to the free-free emitting source inferred by \cite{2017ApJ...842...79M}. The Peanut is significantly closer to $\eta$ Car and smaller than the torus; therefore, it is presumably a younger structure. In principle, its location and shape are consistent with ejecta expelled in an asymmetric --or `one-sided'-- mass ejection. The occurrence of such events in $\eta$ Car was first studied by \cite{2016MNRAS.463..845K}, and later proposed by \cite{2018MNRAS.474.4988S} as a possible explanation for the gap in the CO torus. 

The Peanut is only visible in HCO$^+$ and continuum, without any hints of the other observed molecules. Due to its proximity to $\eta$ Car, it may be directly exposed to intense UV radiation. The survival of dust and molecules at such a short distance of the star requires an efficient shielding mechanism. The wind fossil structures accumulated during multiple cycles could protect the central region as discussed by \cite{2017ApJ...842...79M}. In this regard, the Peanut strikingly resembles the `hook' structure detected towards the NW of $\eta$ Car in forbidden emission lines of [\ion{Fe}{ii}] and [\ion{Fe}{iii}] (see figure 2 in \citealt{2016MNRAS.462.3196G}). The hook exhibits significant changes during the 5.54-year cycle, depending on the ionization state: it is visible in [\ion{Fe}{ii}] across the whole orbit, but fades away in [\ion{Fe}{iii}] during periastron passage, when most of the FUV radiation of $\eta$ Car B is blocked by the primary wind. Provided that the molecular and ionized phases are co-spatial to some extent, then the Peanut is --at least partially-- shielded by this intervening hook, the irradiation of the molecular gas is presumably dominated by MUV photons. However, we must note that survival of HCO$^+$ might not be strongly affected by the ionization state or the shielding efficiency, since this molecule is almost transparent to UV radiation \citep{1995JChPh.103.7006K,2006FaDi..133..231V,2012ApJ...747..114W}. This point is further discussed in Sect. \ref{sec:disc-abu}.

We clearly identify three maxima across the Peanut, as seen in the HCO$^+$ peak intensity map. Two of these maxima are nearly coincident with the brightest peaks in the hook. Indeed, \cite{2016MNRAS.462.3196G} reports that these peaks correspond to the position of Weigelt blobs C and D. This could imply that the Peanut and the blobs are also related, hence confirming a common origin in the eruptive events of the 19th century. We note, though, that the Peanut is slightly \textit{closer}  --in terms of projected distance-- to $\eta$ Car than the hook and the blobs (shifted  towards the SE by about 0.1--0.15 arcsec from peak to peak). This probably indicates that the molecular and ionized gas phases are not totally coupled, as we should expect a spatial transition from ionized to molecular gas (e.g. [\ion{Fe}{iii}] $\rightarrow$ [\ion{Fe}{ii}] $\rightarrow$ HCO$^+$) in the direction of the FUV field.

\subsection{Kinematics of the Peanut}\label{sec:disc-kin}

\begin{figure}
\includegraphics[width=\columnwidth]{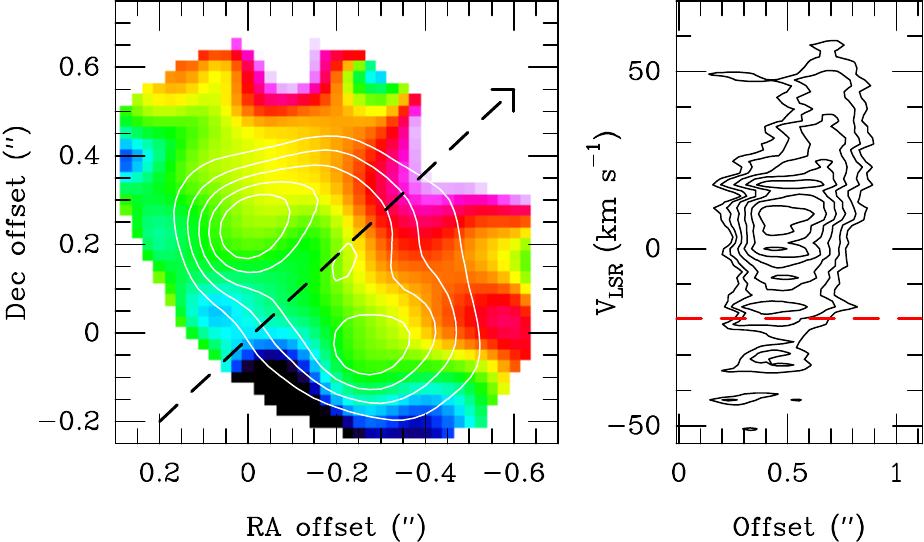}
\caption{Velocity field of the HCO$^+$ gas. Left panel: 1st-order moment map of HCO$^+$ (colour scale), with the integrated intensity superimposed as white contours. Right panel: position-velocity cut in the direction indicated by the black dashed line in the left panel. Contours start at 5$\sigma$ with steps of 3$\sigma$. The red dashed line represents the systemic velocity of $\eta$ Car. }
\label{fig:fig4}
\end{figure}

Analysis of the kinematics of HCO$^+$ could help to understand the origin of this structure and its complex geometry. We found serious discrepancies between the motion of the Peanut and the Weigelt blobs and their associated structures. Firstly, HCO$^+$  is mostly confined to the velocity range ($-100$,$+100$) km s$^{-1}$, with the most intense emission arising from $-20$ to $+70$ km s$^{-1}$. Conversely, the hook is completely blueshifted with respect to the systemic velocity of $\eta$ Car (--19.7 km s$^{-1}$), extending from $-72$ to $-32$ km s$^{-1}$.\footnote{The range reported in \cite{2016MNRAS.462.3196G} is from $-60$ to $-20$ km s$^{-1}$. For $\eta$ Car, $V_\mathrm{LSR} = V_\mathrm{Hel} - 12$ km s$^{-1}$ \citep{2004MNRAS.351L..15S}} Therefore, the HCO$^+$ emission that lies in the velocity range of the hook is almost marginal. As this velocity mismatch suggests, a significant part of the molecular gas might be directly exposed to the FUV field of $\eta$ Car B.

\begin{figure*}
\includegraphics[width=\textwidth]{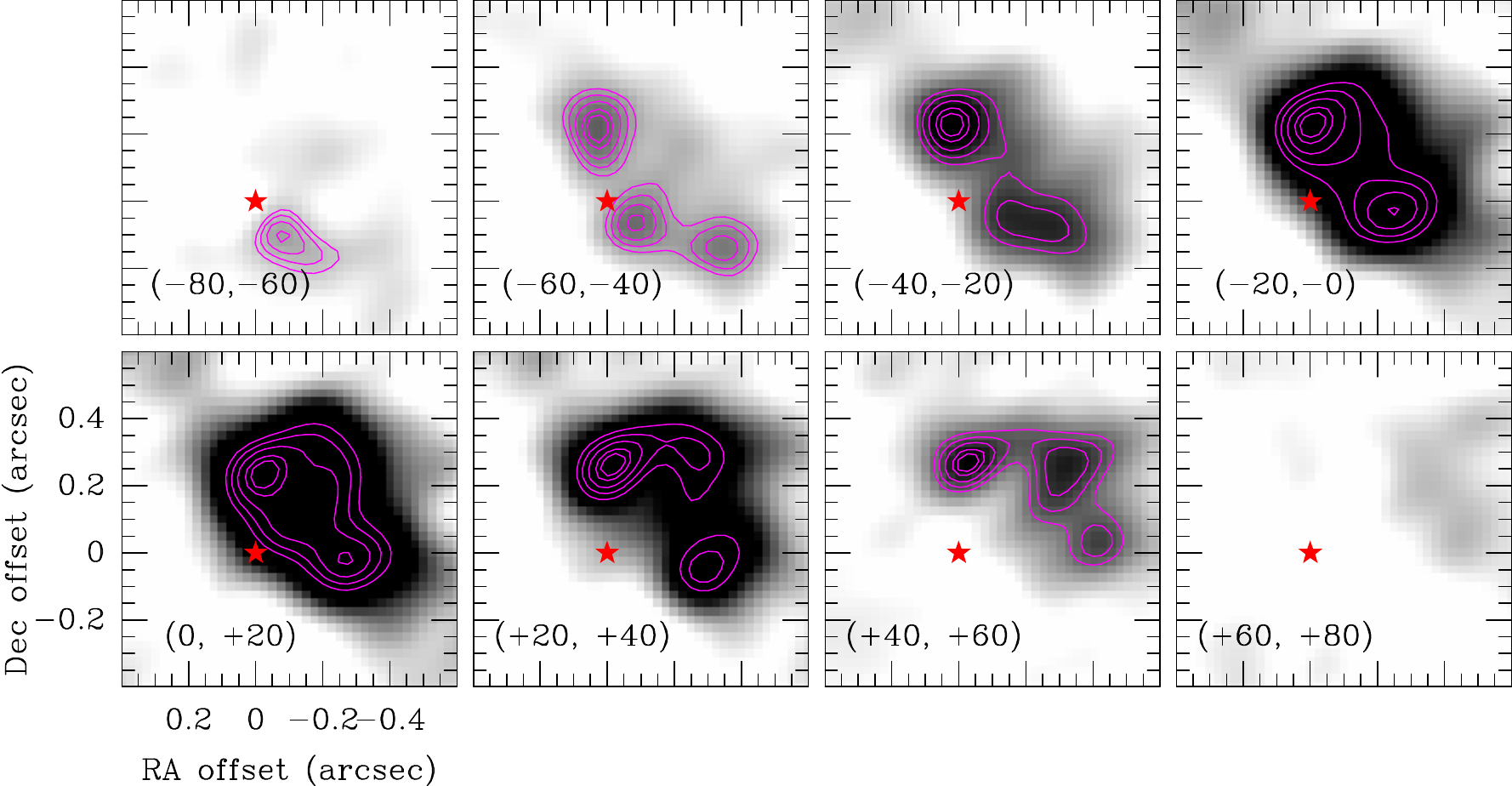}
\caption{Velocity-integrated intensity maps of HCO$^+$ in colour scale, integrated in bins of 20 km s$^{-1}$  from -80 to +80 km s$^{-1}$. All the maps have the same intensity scale. Contours at 60, 70, 80, 90 and 95$\%$ of the peak intensity of each map. The integration range is shown in the bottom left corner of each panel, and the position of $\eta$ Car is indicated by the red marker.}
\label{fig:fig5}
\end{figure*}

Secondly, it is accepted that Weigelt blobs C and D are moving close to the system's orbital plane, which is almost coincident with the Homunculus midplane  \citep{1997ARA&A..35....1D}. We find this motion to be also incompatible with the kinematics of the HCO$^+$: the gas presents a clear gradient from the SE (negative velocities) to the NW (positive velocities), with most of the emission at positive velocities --and therefore redshifted with respect to the systemic velocity of $\eta$ Car-- (see Fig. \ref{fig:fig4}). This velocity distribution, that in principle is consistent with an asymmetric eruptive event like the 1890s outburst, does not match the motion pattern of the Weigelt blobs in the most accepted orbital geometry (e.g. \citealt{2008MNRAS.388L..39O, 2016ApJ...819..131T}), where periastron occurs on the far side of $\eta$ Car A and apastron on the near side. If the Weigelt blobs are escaping from the star towards the NW (i.e. towards apastron) and close to the orbital plane, they are necessarily blueshifted, approaching the observer. Contrarily, the HCO$^+$ gas in this direction is mostly receding.

\begin{figure}
\includegraphics[width=\columnwidth]{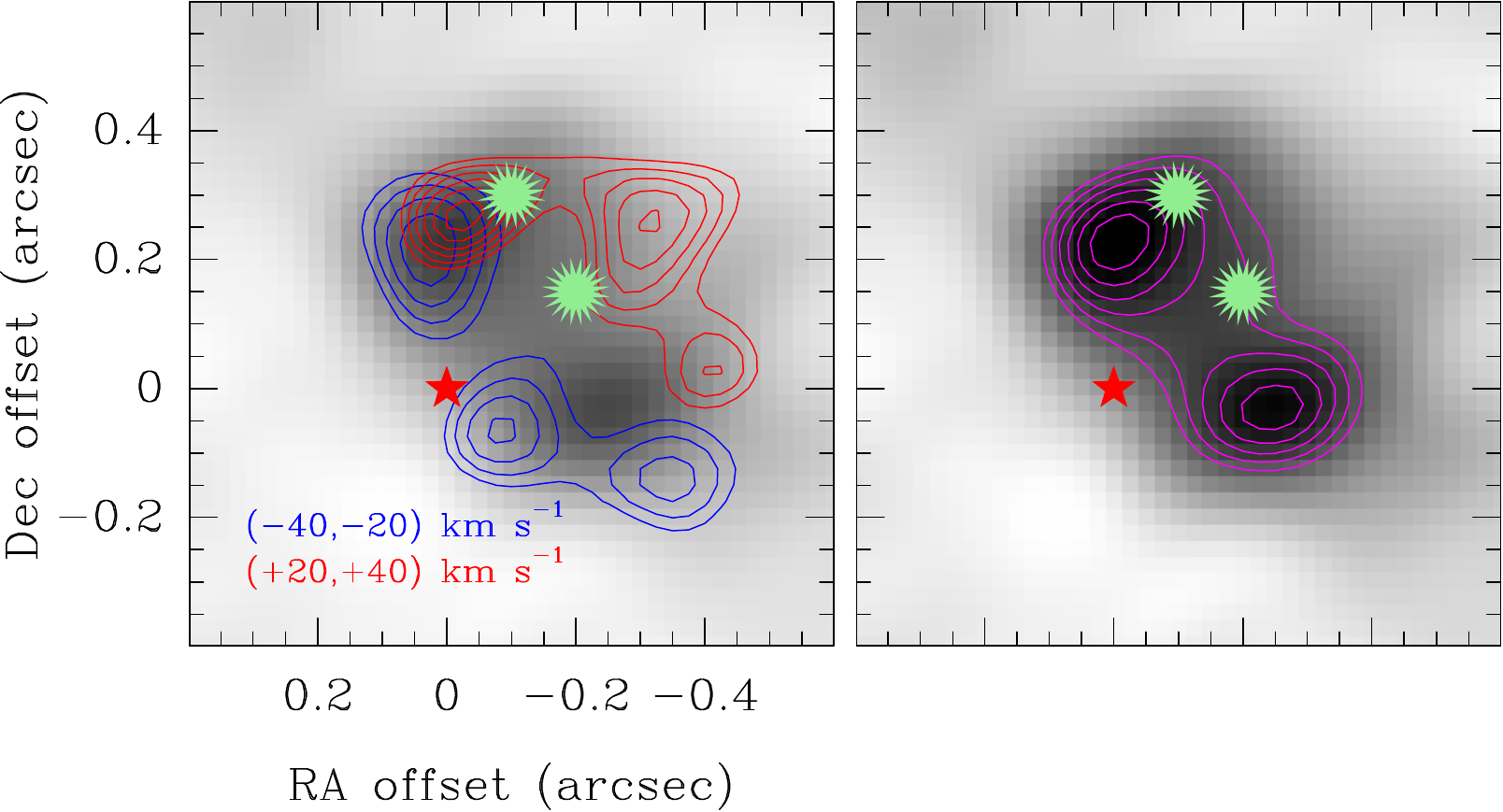}
\caption{Relative position of the Weigelt blobs with respect to the most prominent features of HCO$^+$. Left panel: Integrated line intensity of HCO$^+$ from $-40$ to $-20$ (blue) and from +20 to +40 (red) km s$^{-1}$, as contours, superimposed to the integrated intensity in the range (-20, 0) km s$^{-1}$ in colour scale. Right panel: integrated line intensity of HCO$^+$ from -20 (systemic velocity) to 0  km s$^{-1}$, as contours and colour scale. The position of $\eta$ Car and Weigelt blobs C and D (taken from \citealt{2016MNRAS.462.3196G}) is indicated by the red and green markers, respectively.}
\label{fig:fig6}
\end{figure}

In order to solve this puzzle, a more detailed analysis of the kinematics of the Peanut is required. Fig. \ref{fig:fig5} presents intensity maps of HCO$^+$ integrated in velocity bins of 20 km s$^{-1}$ over the range (--80,+80) km s$^{-1}$. Note that contours are relative to the maximum level of each map to emphasize the peaks. We identify a set of features evolving from SE to NW as we advance towards positive velocities. The most intense emission is concentrated in an irregular bar-like feature in the range (--20, +40) km s$^{-1}$, but two fainter substructures are particularly noteworthy: the two patchy arcs that are clearly visible in the velocity ranges (--60, --40) and (+40, +60) km s$^{-1}$. They are almost symmetrically located with respect to the bulk emission: the blueshifted arc opening towards the NW and the redshifted toward the SE. When taken together, the arcs describe an ellipsoid of $0.6\times0.4$ arcsec with a position-angle of $\sim30$\degr\ east of north, as seen in Fig. \ref{fig:fig6}.

All these features may be explained by a clumpy, roughly conical structure of ejecta from the 1890s outburst. This `cone', with its opening towards apastron, is expanding within the wind-blown cavity described by the colliding wind models proposed by  \cite{2008MNRAS.388L..39O}, \cite{2009MNRAS.396.1308G} and \cite{2013MNRAS.436.3820M} and briefly discussed in Sect. \ref{sec:intro}. In this scenario, the Weigelt blobs would be just UV-illuminated surfaces in this larger structure. Matter within the cavity would be highly ionized during most of the orbital cycle, producing continuum thermal emission \citep{2014ApJ...791...95A}.  The cone would be periodically disrupted as the dense primary wind (slow, but still $\sim$ 5 times faster than the ejecta) flows past during each periastron. Part of the ejecta may be also driven towards the inner walls of the cavity, creating a high-density layer where HCO$^+$ may also arise.

\begin{figure}
\includegraphics[width=\columnwidth]{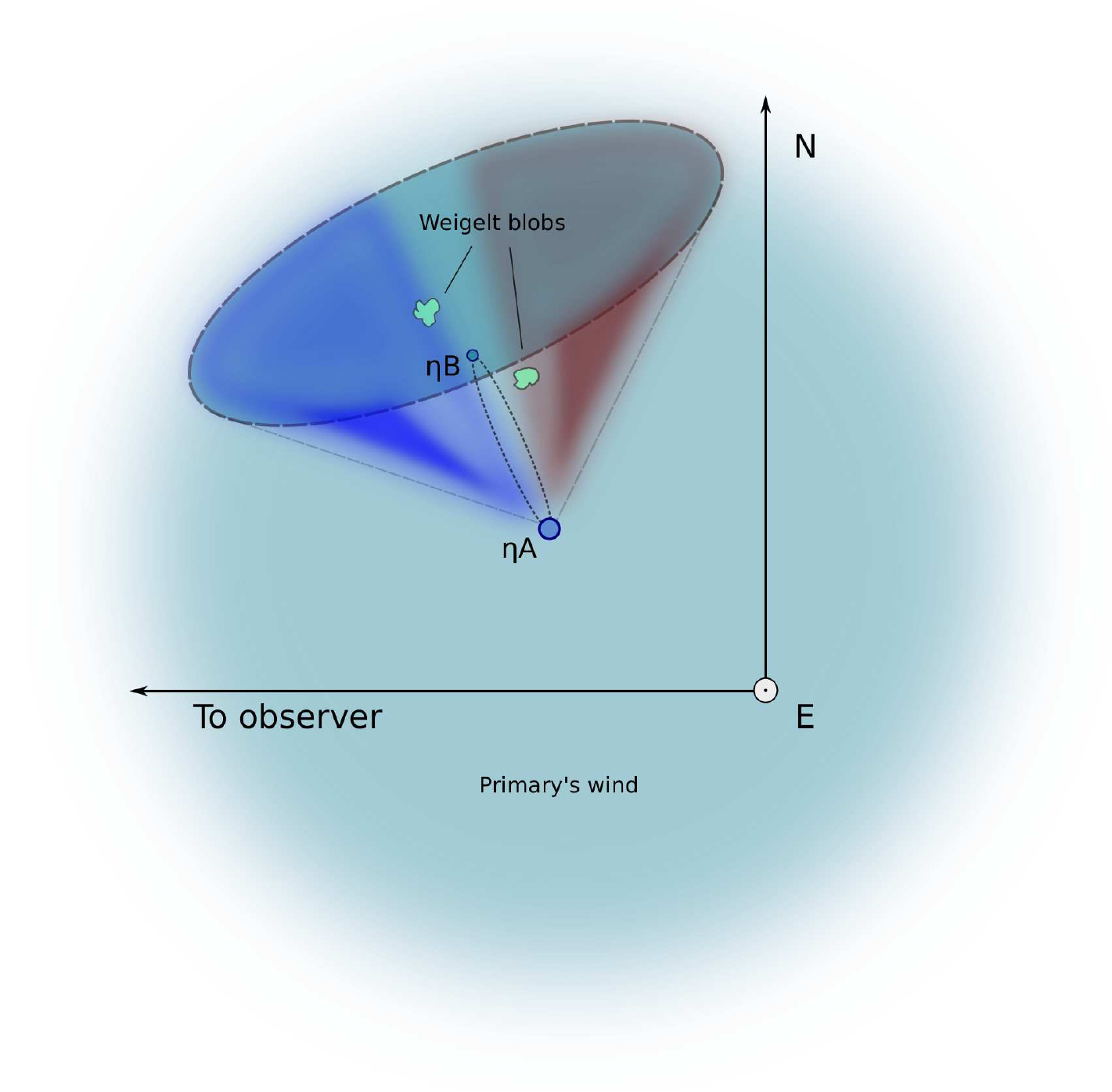}
\caption{Sketch of the viewing geometry of the HCO$^+$ emission. The colours indicate the motion of the gas relative to the observer. The primary wind sphere, the shock cone, the Weigelt blobs and the orbital plane are arbitrarily scaled for illustration purposes.}
\label{fig:geo}
\end{figure}

To put all the pieces together we present a simple sketch model in Fig. \ref{fig:geo}. The proposed orientation of the cone explains the observed velocity pattern of HCO$^+$: the SE side of the cone moves towards the observer (the blueshifted arc) while the NW side moves away (the redshifted arc). The `tangent' parts of the cone (towards NE and SW) would move almost in the plane of the sky, explaining the emission peaks that appear near 0 km s$^{-1}$. These peaks may be caused by an accumulation of material in the line of sight towards these regions. The bulk of the bar would correspond to the densest parts of the structure. This interpretation is able to reconcile the HCO$^+$ gas with the motion of the Weigelt blobs and the ionized Fe structures. Fig. \ref{fig:fig6} shows the position of blobs C and D with respect to the arcs and the bar. The arcs approximately enclose the blobs, which also supports the idea that these structures are moving outwards \textit{within} the cavity created by the colliding winds. The apparent spatial correlation with the HCO$^+$ peaks is then a mere projection effect. The blobs may be actually closer to the apex of the cone and thus directly exposed to a more intense FUV field.

Finally, it is important to note that kinematics of the molecular gas also provide valuable 3D information about the eruption. The location of the Peanut with respect to the star plus the lack of a counterpart for the slowly-moving material in the far side of $\eta$ Car are important evidences that support the asymmetric outburst hypothesis. But most importantly, the remarkable radial velocity dispersion of the HCO$^+$ gas with respect to $\eta$ Car (of about 80--100 km s$^{-1}$) may also indicate a \textit{latitudinal} distribution in the ejecta --in contrast to the equatorial nature of the Weigelt blobs--.

It is important to keep in mind, though, that here we are working at scales comparable to the beam size. To understand gas motions in such small regions and refine the interpretations provided in this work, observations at higher angular resolution are essential. These would provide key insights about the formation and dynamics of molecular gas, its interplay with the ionized and neutral phases, and (especially) its dependency on the orbital phase.

\section{Line absorptions}\label{sec:disc-abs}

\begin{figure*}
\includegraphics[width=\textwidth]{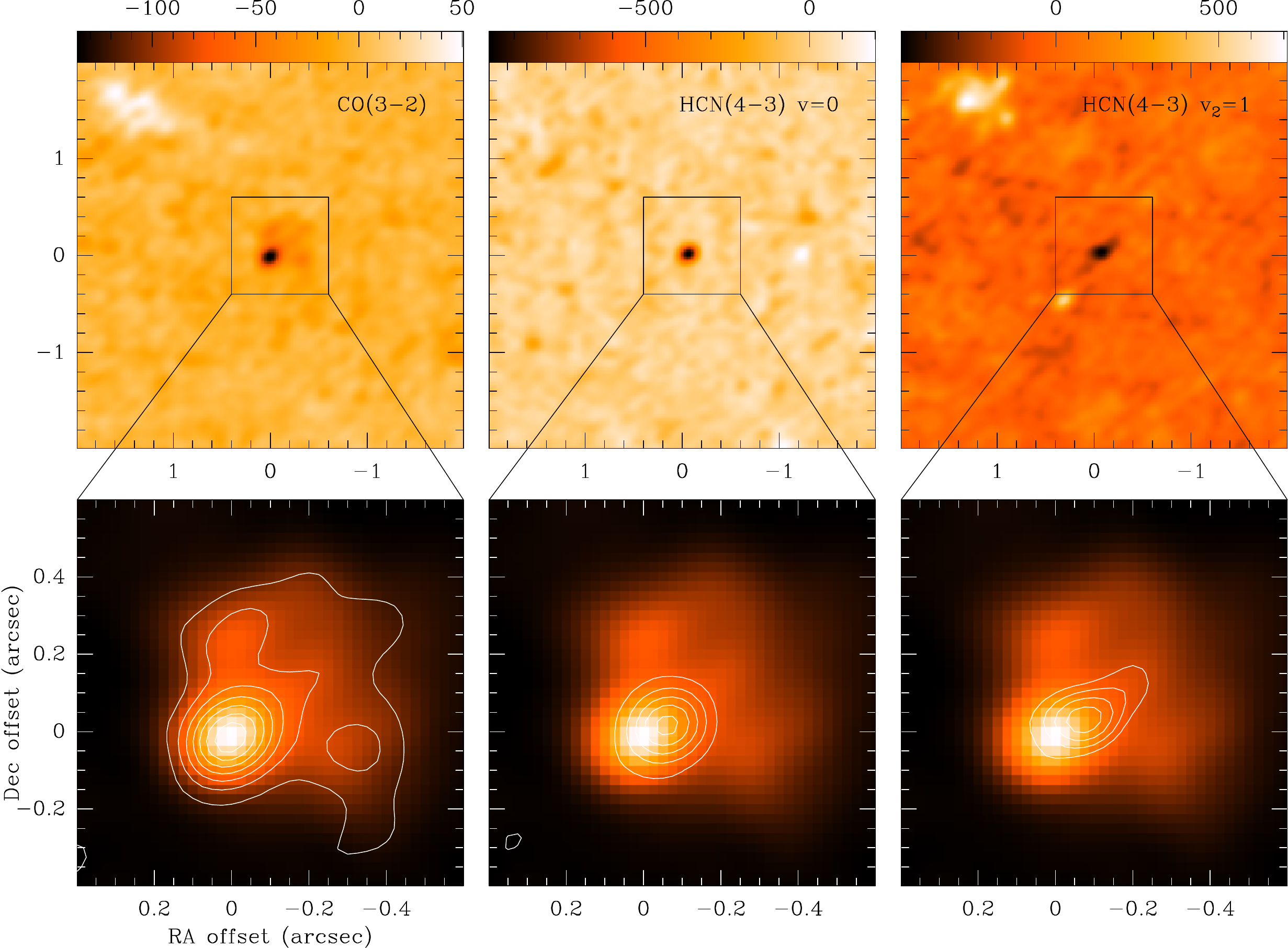}
\caption{Spatial distribution of the absorptions towards $\eta$ Car. Top panels: intensity maps of CO, HCN $v=0$ and HCN $v_2=1$ in colour scale, integrated in the velocity range ($-67$, $-49$) km s$^{-1}$ for a region of 4$\times$4 arcsec. Note that the emission in the top left corner of the CO and HCN $v_2=1$ panels corresponds to CO and HCN $v=0$ emission arising from the torus. Bottom panels: close-in view of the central arcsec for the corresponding lines in the top panels. Integrated intensity as white contours, superimposed to the continuum emission in colour scale.}
\label{fig:abs-distrib}
\end{figure*}

In this section we present a more detailed description of the absorbing gas, whose overall distribution is depicted in Fig.~\ref{fig:abs-distrib}. The three top panels show the distribution in a $4\times4$ arcsec field in false colour; the bottom panels show a zoom in the inner arcsec, with the line absorption as contours superimposed to the continuum. The figure does not include the distribution of the absorbing H$^{13}$CN, which is basically identical to the main isotopologue at both vibrational levels ($v=0$ and $v_2=1$).

Absorbing CO distribution mimics the whole continuum emission, including both the hot star and the Peanut. It is probably a foreground interstellar cloud located somewhere in the foreground of $\eta$ Car. Its velocity ($-$9 km s$^{-1}$) lies halfway between the corresponding velocities of a local cloud (0 km s$^{-1}$) and the Carina arm ($-20$/$-25$ km s$^{-1}$); its line width (0.6 km s$^{-1}$) is typically found in cold clouds. Never the less, this cloud may be related to the dissipating dusty clump proposed by \cite{2019MNRAS.484.1325D} to explain the brightening of the central source, that would act as a `natural coronograph'.

Contrarily, HCN and H$^{13}$CN absorptions are concentrated in a very compact and unresolved region. Its centre is close, but not coincident with the hot star. The projected distance to the star is around 0.08 arcsec (180\,au), which is smaller than the angular resolution but clearly significant after looking the CO absorption at the very centre of the point source. Velocity ($-60$~km~s$^{-1}$) and line widths ($10-12$~km~s$^{-1}$) are similar to those measured in emission in the torus and the Peanut.  All these observational findings are consistent with a bullet-like cloud being expelled from $\eta$~Car at a projected velocity of 40~km~s$^{-1}$. It is finally noteworthy the lack of emission or absorption of the other molecules, which indicates that this hot bullet is likely very N-enriched.

The H$^{13}$CN and HCN lines have similar intensities. This translates, as we develop in Sect. \ref{sec:disc-abu}, to a very low $^{12}$C/$^{13}$C ratio. However, the most striking result is the detection of the vibrationally excited lines with intensities slightly below half of those corresponding to the $v=0$ lines. HCN and H$^{13}$CN  $v_2=1$ excited states correspond to a double degenerate bending mode with an energy of 729.7\,cm$^{-1}$ ($\approx 1050$\,K). It is virtually impossible to populate such energy level exclusively by collisions, and the most probable mechanism is the infrared radiative pumping by the absorption of photons at $\approx 14\mu$m \citep{2015A&A...575A..94B}. The strong infrared continuum of eta Carina ensures the availability of huge amounts of photons for such pumping.

These and other vibrationally excited lines with comparable energy levels have been observed in other astrophysical environments, such as carbon stars (\citealt{1987ApJ...323L..81I, 2001ApJ...549L.125B}), hot cores (\citealt{2013A&A...559A..51E, 2017A&A...604A..32P}), or ultraluminous infrared galaxies (\citealt{2013AJ....146...91I, 2016ApJ...825...44I}). In all those cases, however, the intensity of the vibrationally excited lines remains as a small fraction of the ground level lines. Therefore, the bullet is being shocked and accelerated by a copious amount of infrared photons, in a particularly harsh environment.

\section{Column densities and relative abundances}\label{sec:disc-abu}

\subsection{Column densities}
\begin{figure*}
\includegraphics[width=\textwidth]{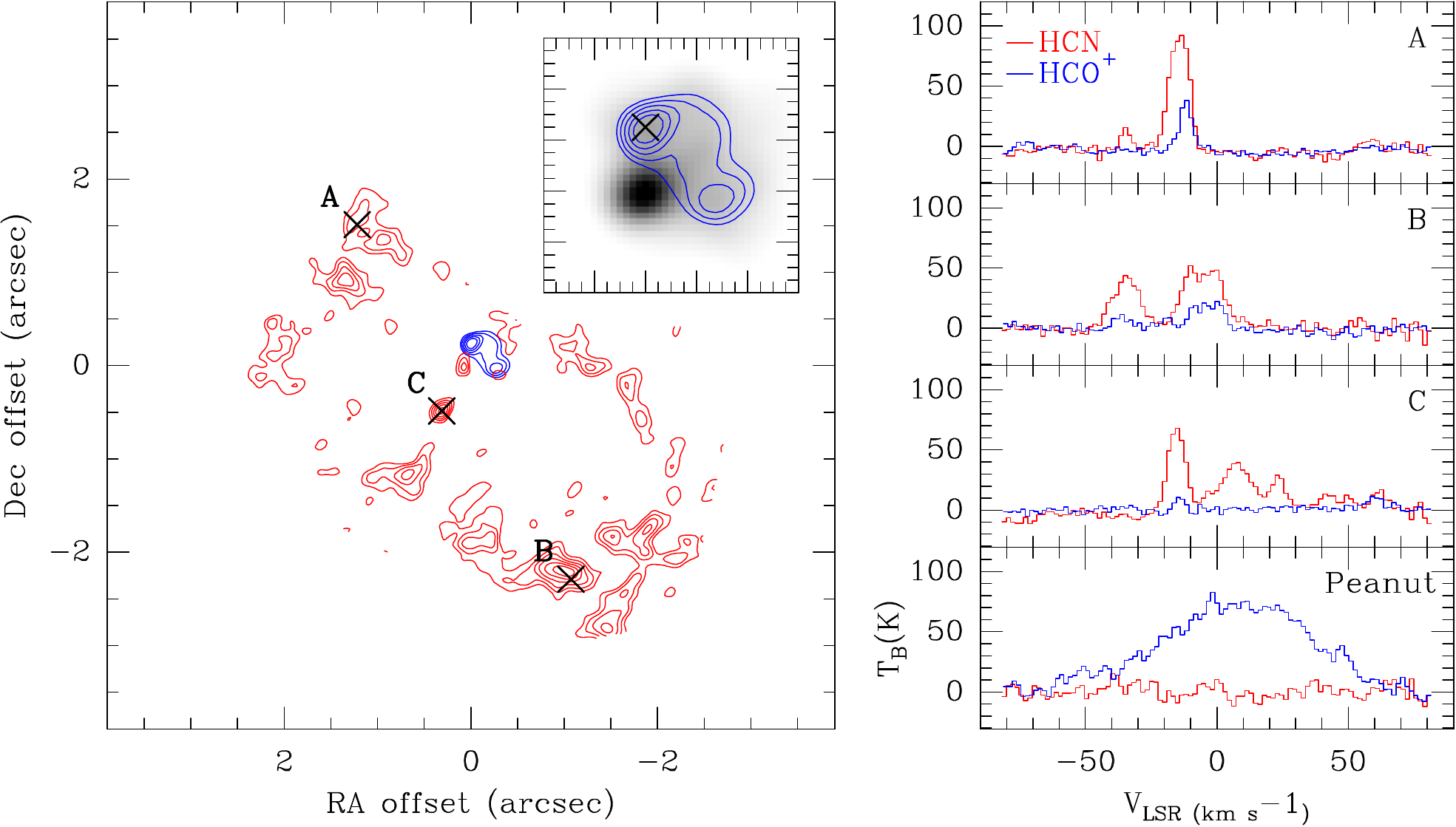}
\caption{Hints of chemical differentiation in the innermost region of the Homunculus. Left panel: line intensity maps of HCO$^{+}$ (blue) and HCN (red) in the velocity range (--100, +80) km s$^{-1}$. The inset shows a close-in view of the Peanut, superimposed to the continuum image (grey scale). Black crosses indicate the positions of the spectra shown to the right. Right panel: HCO$^{+}$ (blue) and HCN (red) spatially averaged spectra in the corresponding positions of the left panel.}
\label{fig:fig3}
\end{figure*}

We estimated the column densities of the observed species in different positions, as depicted in Fig. \ref{fig:fig3}: the torus --positions A and B--, the clump at ($0.\arcsec4,-0.\arcsec65$) --position C--, the Peanut and the absorbing bullet. For the emission lines we assumed that the gas is in local thermodynamic equilibrium (LTE), the emission is optically thin and that radiative excitation dominates over collisions due to the strong radiation field of $\eta$ Car. If gas and dust are thermally coupled, gas temperature may be approximated by the dust temperature. We took the equation by \cite{2003AJ....125.1458S}:

\begin{equation}
T_\mathrm{dust}(\mathrm{K}) \approx 13100 \times D(\mathrm{au})^{-\frac{1}{2}}
\end{equation}

\noindent where $T_\mathrm{dust} $ is the black-body equilibrium temperature of a dust grain at a distance $D$ of the source. Translating angular separation into real distances is almost immediate for the torus after correcting for the inclination (the polar axis of the Homunculus is tilted by 49\degr, \citealt{2006ApJ...644.1151S}). We derive temperatures ranging from 170 to 190 K. For the Peanut, on the contrary, obtaining a proper distance estimate is not straightforward. We note that the maximum at ($0\arcsec,0.\arcsec25$) corresponds to emission present across the whole velocity range (see Fig. \ref{fig:fig5}). This allows us to approximate the real distance by the projected distance. The resulting temperature, of about 500 K, is an upper limit, as we are considering the material that is closest to the star.

For the absorbing bullet, we fully solved the transfer equation with LTE as the only assumption. Considering the proximity to $\eta$ Car, we cannot neglect the excitation term like in the case of absorptions of cold clouds (\citealt{2001A&A...370..576L,  2016PASJ...68....6A, 2018A&A...610A..49L}). The continuum and excitation temperatures result to be 3390 and 500 K, respectively. The four lines are optically thin, with opacities from 0.01 to 0.025. Table \ref{tab:cd} summarizes the resulting column densities and the most relevant abundance ratios. We note that our column densities are consistently lower than those reported by \cite{2012ApJ...749L...4L} by a factor of 3--5. While this may obey to beam filling issues, our estimates are more in line with the column densities derived by \cite{2017ApJ...842...79M} from large beam CO $J=5\rightarrow4$ to $9\rightarrow8$ observations of the Homunculus.

\subsection{Abundances in the torus}
Some molecular abundances vary drastically among different positions. The torus presents [HCO$^+$]/[CO] ratios of $\sim10^{-4}$, comparable to those measured around other massive evolved stars, like the Wolf-Rayet nebula NGC2359  \citep{2003IAUS..212..742R}. Similar ratios are also found in the Orion Bar photodissociation region \citep{2000ApJ...540..886Y} and some molecular clouds, such as TMC-1 \citep{1997ApJ...486..862P}.

On the contrary, we found significantly altered [HCN]/[CO] relative abundances, with values of a few $\sim10^{-3}$. These are almost 100 times larger than the ratios measured in TMC-1 and NGC2359 and just slightly above those of the Orion Bar PDR, but still much lower than in ambients where the grain chemistry dominates \citep{1996A&A...307...52S}. The highest ratios are measured towards the clump and the absorbing bullet, which suggests that these two features may be of similar nature. Interestingly, the [H$^{13}$CN]/[CO] ratio reaches comparable values in all the studied positions.

These high relative abundances of nitrogen-bearing molecules with respect to CO are a typical hint of N-rich gas. The N-enrichment of the Homunculus is a widely studied topic: by means of UV and optical spectroscopy, \cite{1986ApJ...305..867D} demonstrated a N-enrichment factor of 10 with respect to solar abundances; \cite{1997ASPC..120..255D} and \cite{2005ApJ...624..973V} found evidence of heavy C and O depletion, a sign of intensive CNO processing. Still, we are not in a position to report N-enrichment in absolute terms as we lack information about H$_2$ in these specific regions (just overall estimates for the Homunculus). But regardless of the absolute N abundance, the [HCN]/[H$^{13}$CN] ratio is a good proxy to confirm that the torus is made of CNO-processed matter (since $^{13}$C is another intermediate product of the CNO cycle). In every position, we measure [HCN]/[H$^{13}$CN] ratios close to unity. This value is exceptional and holds for the vibrationally-excited lines as well. It indicates an extremely low [$^{12}$C]/[$^{13}$C] isotopic ratio, even for a massive star (typically in the range $3-10$, e.g. \citealt{1984ApJ...284..223L}, \citealt{2019MNRAS.482.1651B}). 

\subsection{Hints of chemical differentiation}

The situation in the Peanut is completely different, as Fig. \ref{fig:fig1} suggested. Towards this region, none of the molecules except HCO$^+$ is detected. Therefore, we constrained the column densities of CO, HCN and H$^{13}$CN taking the 3$\sigma$ level as the line intensity upper limit. In this case, this approach tends to favour these column densities since the maps are noisier towards the continuum emitting region, but we still obtain limits notably below the values measured in the torus. On the other hand, the column density of HCO$^{+}$ is remarkably enhanced, reaching $\sim10^{16}$ cm$^{-2}$. Consequently, the [HCO$^{+}$]/[CO] and [HCO$^{+}$]/[HCN] ratios increase by a factor of up to 125.

The non-detection of CO, HCN and H$^{13}$CN in the Peanut implies that the gas in this region is chemically different from the ejecta of the Great Eruption --i.e. the torus--.  Indeed, \cite{2005A&A...435.1043C} compared the composition of the dust in the Weigelt region and other parts of the butterfly nebula, finding significant differences. Since the ejecta from the 1840s and 1890s eruptions are likely similar in terms of initial CNO abundances, the observed differences may obey to the extreme physical conditions to which the Peanut is exposed.

In dense clouds, one of the principal formation routes of gas-phase HCN is

\begin{equation}
\mathrm{CH_2} + \mathrm{N} \rightarrow  \mathrm{HCN} + \mathrm{H}
\end{equation}

\noindent a neutral-neutral reaction much slower than the ion-molecule reaction that produces HCO$^+$ \citep{2005ApJ...632..302B}. While reaction rates may explain the dominance of HCO$^+$ in the Peanut, by no means it is enough to explain the apparent lack of N-bearing species. Provided that N is abundant to a certain degree --as expected in CNO-processed material-- the non-detection of neither HCN nor H$^{13}$CN may be the consequence of an aggressive photochemistry. We know that the structures traced by [\ion{Fe}{ii}] and [\ion{Fe}{iii}] protect --at least partially-- the Peanut, blocking the most energetic FUV radiation. Because of that, the Peanut is primarily exposed to a substantial flux of MUV photons with wavelengths larger than $1800$ \AA. HCO$^+$ formation and destruction does not directly involve any photochemical route, so this molecule is relatively immune to these photons. On the other hand, HCN is rapidly photodissociated through

\begin{equation}
\mathrm{HCN} + \mathrm{\nu} \rightarrow  \mathrm{CN} + \mathrm{H}
\end{equation}

\noindent producing CN, which in turn is less vulnerable to photodissociation (requiring photons with $\lambda$ < 1100 \AA, \citealt{2006FaDi..133..231V}). If photodissociation is the prevailing destruction mechanism of HCN and H$^{13}$CN in the Peanut, CN should be abundant in this region. In fact, CN was already detected towards $\eta$ Car in single-dish observations by \cite{2012ApJ...749L...4L}, but higher angular resolution is needed to confirm its origin.

Still, the non-detection of CO is challenging and hard to explain, especially when taking into account the high column density of HCO$^+$. The dominant formation pathway of HCO$^+$ in the ISM is the proton transfer reaction

\begin{equation}
 \mathrm{H}_3^+ + \mathrm{CO} \rightarrow \mathrm{HCO}^+ + \mathrm{H}_2
 \end{equation}
 
\noindent (e.g. \citealt{Illies1982}, \citealt{2004A&A...428..117L}), which requires a substantial amount of CO to be efficient. Likewise, HCO$^+$ is mainly destroyed by dissociative recombination through

\begin{equation}
\mathrm{HCO}^+ + \mathrm{e}^- \rightarrow  \mathrm{CO}+ \mathrm{H}
 \end{equation}

\noindent as long as the electron density is high enough  \citep{2000MNRAS.313L..17Y}. This reaction should increase the total CO abundance. The detection of hydrogen RRL emission towards the Peanut somehow supports this mechanism, as the resulting H atoms would be quickly ionized. For that reason, the apparent lack of CO in the Peanut is particularly shocking. However, alternative formation routes of HCO$^+$ that do not require CO cannot be ruled out. Then CO may be simply forming at a much slower rate than HCO$^+$, which would explain its non-detection considering the short time-scales involved ($\sim$ 100 years). In any case, the development of new chemical models is key to understand the chemical evolution of the inner ejecta of $\eta$ Car.

\subsection{Mass estimates}

Determining the masses of the observed molecular structures is not straightforward, since all the abundances refer to CO. It is then fundamental to know the true [CO]/[H$_2$] ratio to translate these results into masses. \cite{2005ASPC..332..294F} and \cite{2006ApJ...645L..41S} provided an average H$_2$ column density through the walls of the Homunculus of around 10$^{22}$ cm$^{-2}$. \cite{2012ApJ...749L...4L} adopted this value to derive a relative abundance of CO of $2.2\times10^{-4}$ for the whole Homunculus.

In principle, we could use this CO abundance, but it may lead to unreliable results, as we are dealing with resolved, clumpy structures, for which the actual H$_2$ column density may be higher. For the torus, a standard [CO]/[H$_2$] abundance of 10$^{-4}$ yields a mass of 0.07$\pm$0.02 $\mathrm{M}_{\sun}$. This value seems exceedingly low considering that around 10--20\% of the Homunculus mass may reside in its equatorial plane \citep{2003AJ....125.1458S}. \cite{2018MNRAS.474.4988S} constrained the mass of the torus to 0.2--1 $\mathrm{M}_{\sun}$ exploring a range of possible abundances. In view of our results, we conclude that CO should be underabundant by a factor of 8--40 in order to be compatible with such mass range. This agrees, again, with the results by \cite{2017ApJ...842...79M}, who reports [CO]/[H$_2$] abundances 10--20 times below cosmic levels.

In the Peanut, the usual approach of determining the H$_2$ mass from CO is not possible due to the non-detection of the latter. HCO$^+$ is the only available H$_2$ tracer, but its relative abundance is also uncertain. HCO$^+$ is found in many astrophysical environments, with abundances ranging from 10$^{-7}$ to 10$^{-10}$ (see e.g. \citealt{1997ApJ...486..316B}, \citealt{2000MNRAS.313..663G}, \citealt{2007AJ....133..364T}). Thus, we could express the mass of the Peanut as a function of the HCO$^+$ abundance, X$_\mathrm{HCO^+}$, by taking a reference value of 10$^{-8}$ (typically measured in oxygen-rich envelopes, \citealt{2011ApJ...743...36P}):

\begin{equation}
M = (0.8\pm0.3) \times \frac{X_\mathrm{HCO^+}}{10^{-8}} \, \mathrm{M}_{\sun}
\end{equation}

Even adopting conservative abundances, such as the value proposed by \cite{2012ApJ...749L...4L} of $5.7\times10^{-8}$, this expression yields masses at least comparable to the mass of ionized gas ejected during the 1890s eruption, which according to \cite{2003AJ....125.3222I} lies in the range 0.1-1 $\mathrm{M}_{\sun}$. None the less, we note that all these estimates must be regarded with caution.

\begin{table*}
\centering
\caption{Column densities and abundance ratios of CO, HCN, H$^{13}$CN and HCO$^{+}$ in the regions depicted in Fig.~2.}
\label{tab:cd}
\begin{tabular}{ccccccccccc}
\hline
Pos. & $T$ & $V_\mathrm{lo}, V_\mathrm{up}$ & $N$(CO) & $N$(HCN) & $N$(H$^{13}$CN) & $N$(HCO$^+$) & $\mathrm{\frac{[HCO^+]}{[CO]}}$ & $\mathrm{\frac{[HCN]}{[CO]}}$ &  $\mathrm{\frac{[HCO^+]}{[HCN]}}$ & $\mathrm{\frac{[HCN]}{[H^{13}CN]}}$ \\
     & (K)   & (km s$^{-1}$) & ($10^{18}$ cm$^{-2})$ & ($10^{15}$ cm$^{-2}$) & ($10^{15}$ cm$^{-2}$) & ($10^{14}$ cm$^{-2}$) & (10$^{-4}$) & (10$^{-3}$) \\
\hline
Tor. A        & 190 & (--35, +0) & $0.73\pm0.01$ & $1.13\pm0.08$ & $1.14\pm0.04$ &$1.48\pm0.14$ & $2.0$ & $1.6$ & 0.13 & 0.99\\
Tor. B        & 170 & (--90, +20) & $1.23\pm0.03$ & $1.72\pm0.14$ & $1.82\pm0.10$ & $4.24\pm0.75$ & $3.5$ & $1.4$ 	& 0.25 & 0.94 \\
Tor. C        & 350 & (-20, +70) & $0.68\pm0.07$ & $3.54\pm0.11$ & $2.68\pm0.10$ & $<4.6$ & $<6.8$ & $5.2$ &$<0.12$ & 1.32 \\
Peanut   & 500 & (--66, +81) & $<0.56$ & $<0.81$ & $<0.87$ & $141.6\pm4.4$ & $>250$ & ... & $>18$ & ...\\
Absorp	 & 500 & (--67, --49) & $<0.28$ &0.62$\pm$0.06	& 0.70$\pm$0.05&  $<3.4$& $...$ & $>2.2$& $<0.55$ &0.89\\
\hline
\end{tabular}

In position Absorp, $N$(HCN) and $N$(H$^{13}$CN) consider the vibrational-ground ($v=0$) and vibrationally-excited ($v_2=1$) states. 
\end{table*}

\section{Concluding remarks}

We analyze ALMA archival observations of continuum, CO, HCN, H$^{13}$CN and HCO$^+$  of the inner 8 arcsec of the Homunculus around $\eta$ Car. The unprecedented resolution of these data allowed us to discover some key ingredients related to the recent mass-loss history of the source and provide complementary insights into the complex wind-wind interactions. The key findings in this work are summarised below:

\begin{enumerate}
\item We reported the detection of the HCN, H$^{13}$CN and HCO$^+$ counterparts of the equatorial torus around $\eta$ Car described by \cite{2018MNRAS.474.4988S} from CO $J=2\rightarrow1$ observations. In addition, the CO $J=3\rightarrow2$ map presented here improves resolution by a factor of $\sim$5, confirming the clumpy nature of this structure.
\item We reported the detection of an extended, asymmetric structure in the inner arcsec, the Peanut, located NW of the star and only visible in continuum and HCO$^+$.
\item We studied the morphology and kinematics of the Peanut in the context of previously known fossil wind structures and ejecta from the 19th century eruptions. We found several velocity features that allowed us to identify the Peanut as a roughly conical structure of ejecta expanding outwards within the cavity blown-out by the current wind interactions. The ejecta, rich in dust and gas, encompasses the Weigelt blobs, which may be partially ionized surfaces of the structure, exposed to intense MUV/FUV fields. This scenario is consistent with the current knowledge of the inner ejecta of $\eta$ Car and existing SPH models.
\item We found a strong absorption feature almost coincident with the star, only visible in HCN and H$^{13}$CN. We also stress the detection of the corresponding vibrationally-excited states of these lines, with unusually high relative intensities. The high temperatures required to excite these states prove that the absorbing gas is close to the photosphere rather than part of a foreground cloud (i.e. the obscuring debris proposed by \citealt{2019MNRAS.484.1325D} to explain the light curve.)
\item We analysed the chemistry of the torus and the Peanut, finding remarkably different molecular abundances. We also compared the results with other astrophysical environments. The abundance of N-bearing species with respect to CO and the extremely low [HCN]/[H$^{13}$CN] ratios measured in the torus are consistent with CNO processed matter. Contrarily, the Peanut is very bright in HCO$^+$ but it is not detected in any of the other lines. The non-detection of CO, the most ubiquitous molecule after H$_2$, is exceptionally intriguing, as it is generally involved in the formation and destruction of HCO$^+$.
\end{enumerate}

All of these findings uncover a complex and challenging scenario. The Peanut is an environment subject to violent changes, where very energetic processes act upon the products of the CNO cycle in very short time scales. Under these extreme conditions, the constituents of molecular gas are exposed to intense UV fields and may have no time to reach chemical equilibrium. This would explain the non-detection of HCN and H$^{13}$CN towards this region. However, if further surveys confirm that the Peanut is indeed deficient in N-bearing species, other alternative scenarios may come into play: for instance, the Peanut may be ejecta from $\eta$ Car B.

Molecular spectroscopy has proven to be a valuable tool, providing a new, complementary approach to the physics, chemistry, and evolution of $\eta$ Car and its surroundings. Monitoring of the chemical evolution of the Peanut, together with a revision of stellar evolution models, is of paramount importance to explain the measured abundances. Moreover, observations at different orbital phases will be key to understand the dynamical evolution and excitation mechanisms of the observed ejecta structures. Finally, extending the observations to other molecules and transitions with the highest resolution possible will allow to constrain the physical conditions of the gas and complete the chemical puzzle of $\eta$ Car.


\section*{Acknowledgements}

We thank the anonymous referee for the insightful comments provided. This paper makes use of the following ALMA data: ADS/JAO.ALMA\#2016.1.00585.S. ALMA is a partnership of ESO (representing its member states), NSF (USA) and NINS (Japan), together with NRC (Canada), MOST and ASIAA (Taiwan), and KASI (Republic of Korea), in cooperation with the Republic of Chile. The Joint ALMA Observatory is operated by ESO, AUI/NRAO and NAOJ. J.R.R. acknowledges the support from projects ESP2015-65597-C4-1-R and ESP2017-86582-C4-1-R (Ministerio de Ciencia, Innovaci\'on y Universidades).

\bibliographystyle{mnras}
\bibliography{references} 

\bsp	
\label{lastpage}
\end{document}